\documentclass[a4paper,12pt,leqno]{article}

\usepackage{graphicx}
\usepackage{subeqn}
\usepackage{amssymb}
\usepackage{lineno}
\usepackage{bm}
\usepackage{array}
\usepackage{color}
\usepackage{subfigure}

\date{ }

\begin{document}

\title{A Novel Exact Representation of Stationary Colored Gaussian Processes \\
\small Fractional Differential Approach
\footnote{Publication info: Cottone G., Di Paola M., R. Santoro, A Novel Exact Representation of Stationary Colored Gaussian Processes (Fractional Differential Approach), Journal of Physics A: Mathematical and Theoretical, 2010, Volume 43, Number 8
page 085002,
doi:10.1088/1751-8113/43/8/085002} }

\author{Giulio Cottone $^{1,2}$ \footnote{E-mail: giulio.cottone@tum.de; giuliocottone@yahoo.it}, Mario Di Paola $^{1}$ and Roberta Santoro $^1$\\
\small $^1$ Dipartimento di Ingegneria Civile, Aerospaziale ed Ambientale, \\ 
\small Universit\'{a} degli Studi di Palermo, Viale delle Scienze, 90128 Palermo, Italy\\
\small $^2$ Engineering Risk Analysis Group, Technische Universit\"at M\"unchen,\\
\small Theresienstr.90, building N6, 80290, Germany
}


\maketitle


\begin{abstract}
A novel representation of functions, called generalized Taylor form, is applied to the filtering of white noise processes. It is shown that every Gaussian colored noise can be expressed as the output of a set of linear \textit{fractional stochastic differential equation} whose solution is a weighted sum of fractional Brownian motions. The exact form of the weighting coefficients is given and it is shown that it is related to the fractional moments of the target spectral density of the colored noise.
\end{abstract}

\section{Introduction}
\noindent 
In this paper a novel procedure to represent a stationary Gaussian process by filtering a Gaussian white noise process is reported. 

Linear stochastic differential equation excited by a Gaussian white noise process acts as a filter, returning an output that is a Gaussian process with Power Spectral Density (PSD), or correlation function, related to the equation's parameters. The system linearity, of course, implies the normality of the output process. 
Depending on the structure and on the coefficients of the filter equation, processes with different PSD might, at least theoretically, be obtained. In practice, such a task is not simple at all, and many papers in literature deal with the characterization of filter in order to fit some target PSD. 

Such a spread interest depends on the fact that many real phenomena of engineering and physical interest are indeed modeled as stationary Gaussian processes with PSD that is known from experimental works. To cite just fews, the most used in the fields of earthquake, wind and ocean engineering are the Tajimi-Kanai \cite{taji60}, the Davenport and the Kaimal \cite{dave61}, \cite{kaim72} and the Pierson-Moskowitz \cite{pier69} spectra, respectively.
Representation of such processes as output to linear differential equations excited by white noises is a key issue in dynamical analysis of single and multi degree of freedom systems. In earthquake engineering readers can find some example in \cite{clou75}, \cite{fals99}, \cite{fals00} while in the context of offshore design, shaping filters for random response analysis are studied in \cite{tham92}.
Auto Regressive (\textit{AR}), Moving Average (\textit{MA}) and their combination (\textit{ARMA}) models have been extensively used by many authors to represent colored processes. In \cite{span83} the applicability of \textit{AR}, \textit{MA} and \textit{ARMA} algorithms to represent the wave motion following a Pierson-Moskowitz spectrum is discussed. Further, in \cite{span86} an analog filter approximation is presented for the Jonswap spectrum. 

In a spectral fitting problem, a common issue of such spectrum approximations is that the determination of filter parameters defining the PSD digital model, relies on some optimization criterion. Indeed, once the \textit{AR}, \textit{MA} or \textit{ARMA} model is selected, the coefficients defining the pulse transfer function that must fit the target PSD lack of any further interpretation.

Recently, the first two authors introduced the fractional calculus and the generalized Taylor expansion involving Fractional Spectral Moments (FSM) to describe both PSD and correlation function in the whole domains $]-\infty <\omega< \infty[$ and $]-\infty <\tau< \infty[$. That is, the spectral fitting problem is easily recast in finding integrals of the target function, which have the meaning of complex moments, and represent the coefficients of a series representation. This approach based on fractional calculus is of great generality and has also been presented for density estimation in probability in \cite{cott09} and \cite{cott10}.

The new issue presented in this paper, extensively based on the latter concepts, is to find out the differential equation coupled to the spectral representation based on complex moments. In particular, it will be assumed to deal with some spectral data coming from experiments, i.e. the target PSD. Then, by the fractional spectral moments two results will be presented.

The first result is a representation of the stationary Gaussian process. Indeed, it is shown that a process with assigned target density can be represented once the FSM are known, by means of an expression that involves fractional derivative of a Gaussian noise. Such processes have been recently shown to be fractional Brownian motions (fBm) \cite{orti08}. Then, every Gaussian colored noise can be thought as superpositions of weighted fBm, and the weights are determined by the FSM.

From this result a further second very remarkable result is achieved. Indeed, by proper application of the fractional calculus, a \textit{linear fractional stochastic differential equation}, whose solution is the colored noise process with target density, is found.

\section{A new representation formula for Fourier pair functions}

In this section, some preliminary concepts and definitions on fractional operators are summarized for clarity's sake as well as to introduce appropriate notations.

Let us consider a Fourier transformable function $f(t)$ and let us denote $\varphi(\omega)$ its Fourier transform,that is
\begin{equation}
\varphi \left( \omega  \right) = {\cal F}\left\{ {f\left( t \right);\omega } \right\} = \int_{ - \infty }^\infty  {f\left( t \right)e^{i\omega t} {\rm{d}}t} 
\label{eq1}
\end{equation}
and its inverse transform is written as
\begin{equation}
f\left( t \right) = {\cal F}^{ - {\rm{1}}} \left\{ {\varphi \left( \omega  \right);t} \right\} = \frac{1}{{2\pi }}\int_{ - \infty }^\infty  {\varphi \left( \omega  \right)e^{ - i\omega t} {\rm{d}}\omega } 
\label{eq2}
\end{equation}

Let us recall the definitions of the Riesz fractional integrals and derivatives as follows
\begin{subequations}\label{eq4}
\begin{eqnarray}\label{eq4a}
\left( {I^\gamma  f} \right)\left( t \right) = \frac{1}{{2\nu \left( \gamma  \right)}}\int_{ - \infty }^\infty  {\frac{{f\left( \tau  \right)}}{{\left| {t - \tau } \right|^{1 - \gamma } }}{\rm{d}}\tau } ;\,\,\,\,\,\,\,\,\rho > 0,\,\rho  \ne 1,3,5,...
\end{eqnarray}
\begin{eqnarray}\label{eq4b}
\left( {{\cal{D}}^\gamma  f} \right)\left( t \right) = \frac{1}{{2\nu \left( -\gamma  \right)}}\int_{ - \infty }^\infty  {\frac{{f\left( {t - \tau } \right) - f\left( t  \right)}}{{\left| {\tau } \right|^{1 + \gamma } }}{\rm{d}}\tau }
\end{eqnarray}
\end{subequations}
where $\nu(\gamma)=\Gamma(\gamma)\rm{cos}(\gamma \pi/2)$ with $\Gamma \left(  \bullet  \right)$ is the Euler gamma function and $\gamma = \rho +\rm{i}\eta, \rho >0, \eta \in \mathbb{R}$. 
Their Fourier transforms, in case $0<\rho<1$, (\cite{samk93}, p.217) are
\begin{subequations}\label{qq5}
\begin{eqnarray}\label{qq5c}
{\cal F}\left\{ {\left( {I_{}^\gamma  f} \right)\left( t \right);\omega } \right\} = \left| \omega  \right|^{ - \gamma } {\cal F}\left\{ {f\left( t \right);\omega } \right\}
\end{eqnarray}
\begin{eqnarray}\label{qq5d}
{\cal F}\left\{ {\left( {{\cal{D}}^\gamma  f} \right)\left( t \right);\omega } \right\} = \left| \omega  \right|^\gamma  {\cal F}\left\{ {f\left( t \right);\omega } \right\}
\end{eqnarray}
\end{subequations}

Comparing eqs.(\ref{qq5c})-(\ref{qq5d}) it can be simply verified that the relation
$\left( {{\cal{D}}^\gamma  f} \right)\left( t \right) \equiv \left( {I^{-\gamma}  f} \right)\left( t \right)$ holds true. Readers should keep in mind that such condition is valid at least for Fourier transformable functions here considered. 
This property allows to calculate the fractional integral in eq.(\ref{eq4a}) by simply changing the sign inside the definition of eq.(\ref{eq4b}) and vice-versa. It is to be stressed that this fact is not trivial at all. Indeed, dealing with functions that are not Fourier transformable, in general $\left( {{\cal{D}}^\gamma  f} \right)\left( t \right) \ne \left( {I^{ - \gamma } f} \right)\left( t \right)$ (see \cite{samk93}, p.214 and p.112-3, Lemma 5.2). 

For brevity's sake, definitions of the other fractional operators as Riemann-Liouville (RL) fractional integral and derivative and of the Marchaud fractional derivative, with their Fourier transforms are provided in Appendix A1. Readers can find a complete theory on such operators in the excellent monograph of Samko et al. \cite{samk93}.

For functions that are Fourier pairs, as those considered in this paper, the eqs.(\ref{qq5}) are very useful to calculate the fractional operators in an easier way with respect to definitions in eqs.
(\ref{eq4}). Indeed, by applying inverse Fourier transform to eqs.(\ref{qq5}), it leads to
\begin{equation}
\left( {I_{}^\gamma  f} \right)\left( t \right) = ({\cal{D}}^{-\gamma} f)(t)=\frac{1}{{2\pi }}\int_{ - \infty }^\infty  {\left| \omega  \right|^{ - \gamma } \varphi \left( \omega  \right)e^{ - i\omega t} {\rm{d}}\omega } 
\label{qq6b}
\end{equation}

From the above equation it can be observed that the convolution integrals (\ref{eq4a}) and (\ref{eq4b}) may be evaluated by making firstly the Fourier transform of $f(t)$, namely $\varphi \left( \omega  \right)$  and then making the Fourier transform of 
$\left| \omega  \right|^{ - \gamma } \varphi \left( \omega  \right)$. This remark is very suitable especially when the Fourier transformation of $f(t)$ is known in analytical form.

Eq.(\ref{qq6b}) evaluated in $t=0$ assumes the particular meaning of fractional moments of $\varphi(\omega)$ in the form
\begin{equation}
\mu \left( { - \gamma } \right)\mathop  = \limits^{def} 2\pi \left( {I_{}^\gamma  f} \right)\left( 0 \right) = \int_{ - \infty }^\infty  {\left| \omega  \right|^{ - \gamma } \varphi \left( \omega  \right){\mathop{\rm d}\nolimits} \omega } 
\label{qq8}
\end{equation}

It has been recently shown \cite{cott09}, that the quantities in 
eq.(\ref{qq8}) are able to represent both the function $f(t)$ and $\varphi(\omega)$ for 
symmetric real function. 
Since the goal of the paper is to represent univariate processes as output of filtered white noise and the Gaussian process is fully characterized in the probabilistic setting by the (symmetric) PSD, in the following, for readability's sake we suppose that $f(t)$ is symmetric. This leads to simplifications, as the Riesz integral definition in eq.(\ref{eq4}) can be rewritten in the form
\begin{equation}
\left( {I^\gamma  f} \right)\left( 0 \right) = \frac{1}{{\nu \left( \gamma  \right)}}\int_0^\infty  {\tau ^{\gamma  - 1} f\left( \tau  \right)} {\mathop{\rm d}\nolimits} \tau 
\label{qq12}
\end{equation}
and directly interpreted as Mellin transform (see Appendix A2). 
In this paper we will deal exclusively with $f(t) \in \mathbb{R}$ and symmetric and consequently  $\varphi(\omega) \in \mathbb{R}$ is also symmetric. Extension to more general conditions is given in Appendix A1. Then, the corresponding representations of the symmetric functions $f(t)$ and $\varphi(\omega)$ are given by
\begin{subequations}\label{qq13}
\begin{eqnarray}\label{qq13a}
f\left( t \right) = \frac{1}{{2\pi i}}\int_{\rho  - i\infty }^{\rho  + i\infty } {\frac{{\nu(\gamma)}}{{2\pi }}\mu } \left( { - \gamma } \right)\left| t \right|^{ - \gamma } {\rm d} \gamma 
\end{eqnarray}
\begin{eqnarray}\label{qq13b}
\varphi \left( \omega  \right) = \frac{1}{{4\pi i}}\int_{\rho  - i\infty }^{\rho  + i\infty } {\mu \left( { - \gamma } \right)} \left| \omega  \right|^{\gamma  - 1} {\rm{d}}\gamma 
\end{eqnarray}
\end{subequations}

Both the integrals are performed along the imaginary axis with fixed real part $\rho$ belonging to the so-called \textit{fundamental strip} of the Mellin transform. A deeper insight on Mellin transform related concepts and more information on the application to fractional derivatives, here omitted for readability's sake, are reported in Appendices A1 and A2.  

Keeping in mind eq.(\ref{qq8}), eqs.(\ref{qq13}) can be understood as a Taylor integral expansion, because by the knowledge of the fractional integrals (or derivatives) in $t=0$, it may fully reconstruct the function in the whole domain. For this reason, eqs.(\ref{qq13}) will be indicated as \textit{generalized Taylor integral forms}. 
This particular form involving fractional moments of Fourier transform of $f(t)$ is entirely new at our best knowledge. It is to be remarked the $\mu (-\gamma)$ is able to represent both $f(t)$ and its Fourier transform in the whole domains.

Eq.(\ref{qq8}) assigns further a geometrical and physical meaning to fractional operators in the Taylor integral form, relating them with the concept of complex moments of the function $\varphi(\omega)$. Applications of eqs.(\ref{qq13}) for the case of correlation functions and PSD may be found in \cite{cott09b}, where the fractional moments of the power spectral density assume the meaning of \textit{fractional spectral moments} and were labeled by $\Lambda (\gamma)$.

It is important to observe that in the generalized Taylor expansions (\ref{qq13}), the integral is performed along the imaginary axis and, under the hypothesis that the direct Mellin transform in eq.(\ref{qq8}) exists, such integrals do not diverge in virtue of the Mellin inverse theorem. Moreover it has to be stressed that the integration does not depend on the particular choice of $\rho$ except for the limitation, as previously outlined, that $\rho$ shall belong to the fundamental strip of the Mellin transform. This may be explained by the fact that the integrand in the fundamental strip is holomorphic. More information on this topic may be found in \cite{cott09}.

Hereinafter, taking full advantage of the results presented in this section, the fractional filter representing a given target PSD will be found.

\section{Transfer function representation by H-Fractional Spectral Moments}
\label{sec3}
Objective of this paper is to represent a normal stationary process with assigned power spectral density as the output of a fractional differential equation. To this aim, let us consider a Gaussian white noise process, $W(t)$ with zero mean and correlation function $E[W(t)W(t+\tau)]=q\;\delta(t) $, and power spectral density $S_W=q/(2\pi)$, where $q$ is the intensity parameter.
Let us indicate the ideal linear system as
\begin{equation}
{\cal{L}}(Y(t))=W(t)
\label{qq14}
\end{equation}
where ${\cal{L}}\left(  \bullet  \right)$ is a linear differential operator applied to the response $Y(t)$. The solution to eq.(\ref{qq14}) can be characterized by the impulse response function $h(t)$ and its Fourier transform $H(\omega)$ namely the transfer function, and can be expressed by Duhamel integral
\begin{equation}
Y\left( t \right) = \int_{ - \infty }^t  {h\left( {t - \tau } \right)W\left( \tau  \right){\rm{d}}\tau } 
\label{qq15}
\end{equation}

Indicating by $S_Y (\omega)$ and $S_W (\omega)$ the power spectral density of the output and the input, respectively, from the linearity of the system it follows 
\begin{equation}
S_Y \left( \omega  \right) = \left| {H\left( \omega  \right)} \right|^2 S_W \left( \omega  \right) = \frac{q}{{2\pi }}\left| {H\left( \omega  \right)} \right|^2 
\label{qq16}
\end{equation}
where $S_Y (\omega)$ is the target spectrum.
Eq.(\ref{qq14}) may be considered as a filter of the white noise process. Let us now suppose that the differential operator ${\cal{L}}\left(  \bullet  \right)$ is such that $S_Y (\omega)$ overlaps an assigned PSD arising from a physical phenomenon. In order to find the unknown operator ${\cal{L}}\left(  \bullet  \right)$ we assume $ Arg\left( {H\left( \omega  \right)} \right) = 0$, that is, from eq.(\ref{qq16}) the transfer function can be expressed as
\begin{equation}
H\left( \omega  \right) = |H(\omega)| = \sqrt {\frac{{2\pi }}{q}S_Y \left( \omega  \right)} 
\label{qq17}
\end{equation}

Correspondingly, as the transfer function is assumed to be real, the impulse response function of the linear system ${\cal{L}}(Y(t))$ remains symmetric. That is, by enforcing the condition ${\mathop{\rm Im}\nolimits} \left( {H\left( \omega  \right)} \right)=0$ we get a non causal differential equation. In spite of the causality condition is violated, the output of eq.(\ref{qq14}) remains a strictly stationary Gaussian process.
In the following, we will firstly introduce the fractional moments of the function $H(\omega)$; then we will represent $H(\omega)$ as sum of fractional moments and finally the expression of the process $Y(t)$ with target PSD $S_Y (\omega)$ will be given in terms of the fractional moments of $H(\omega)$. As first step, in order to represent the transfer function $H(\omega)$, let us define the fractional moment of $H(\omega)$ labeled in the following as ${\Pi _H \left( { - \gamma }\right)} \in \mathbb{C}$, that in virtue of eq.(\ref{qq8}) is written as
\begin{eqnarray}\label{qq18}
\Pi _H \left( { - \gamma } \right)\mathop  = \limits^{def} \int_{ - \infty }^\infty  {\left| \omega  \right|^{ - \gamma } H\left( \omega  \right){\rm d} \omega ,\,\,\,\,\,\,\,\,\,\,{\rm Re} \gamma  > 0} 
\end{eqnarray}
that will be called \textit{H-Fractional Spectral Moment} (H-FSM) function. By means of eq.(\ref{qq13}), the H-FSM function allows us to fully reconstruct both the function $H(\omega)$ and $h(t)$ in the whole domains. The H-FSM are fractional integrals and derivatives of $h(t)$ evaluated in zero (as given in eq.(\ref{qq8})), that is
\begin{eqnarray}\label{qq19}
2\pi \left( {I^\gamma  h} \right)\left( 0 \right) = 2\pi ({\cal{D}}^{-\gamma} h)(0)= \Pi _H \left( { - \gamma } \right),\,\,\,\,\,\,\,\,\,\,\,\,\,\,\,\,\,\,\,\,\,{\rm Re} \gamma  > 0
\end{eqnarray}

Specifying the eq.(\ref{qq13}), previously written for two general Fourier pair, we obtain the representation of the impulse response in the time domain as
\begin{equation}
h\left( t \right) = \frac{1}{{\left( {2\pi } \right)^2 i}}\int_{\rho  - i\infty }^{\rho  + i\infty } {\nu \left( \gamma  \right)\Pi _H \left( { - \gamma } \right)t^{ - \gamma } {\rm d} \gamma ,\,\,\,\,\,\,\,\,\,\,\,\,\,t > 0,\,\,\,\,0 < {\rm Re} \gamma  < 1} 
\label{qq20}
\end{equation}
and the transfer function in the form
\begin{equation}
H\left( \omega  \right) = \frac{1}{{4\pi i}}\int_{\rho  - i\infty }^{\rho  + i\infty } {\Pi _H \left( { - \gamma } \right)\left| \omega  \right|^{\gamma  - 1} {\rm d} \gamma ,\,\,\,\,\,\,\,\,\,\,\,\,\,\,\,\,\,\,\,\,\,0 < {\rm Re} \gamma  < 1} 
\label{qq21}
\end{equation}

It might be useful to consider the H-FSM as a third representation in the $\gamma$ variable of the response of the dynamical system under study.

The integrals involved in (\ref{qq20}) and (\ref{qq21}) can be approximated in discrete form, operating a truncation in the $\eta$ axis. Indeed, calculating the integral up to a certain value, called ${\bar \eta }$, dividing $\left[ { - \bar \eta ,\bar \eta } \right]$ in $2m$ intervals of amplitude $\Delta \eta  = \bar \eta /m$, with $m \in \mathbb{N}$ and evaluating the integrals at the values $\gamma_k=\rho + {\rm{i}} k \Delta\eta$, eqs.(\ref{qq20}) and (\ref{qq21}) can be approximated in the form
\begin{eqnarray}\label{qq22}
h\left( t \right) \cong \frac{{\Delta \eta }}{{\left( {2\pi } \right)^2 }}\sum\limits_{k =  - m}^m {\nu \left( {\gamma _k } \right)\Pi _H \left( { - \gamma _k } \right)t^{ - \gamma _k } } 
\end{eqnarray}
\begin{eqnarray}\label{qq23}
H\left( \omega  \right) \cong \frac{{\Delta \eta }}{{4\pi }}\sum\limits_{k =  - m}^m {\Pi _H \left( { - \gamma _k } \right)\left| \omega  \right|^{\gamma _k  - 1} } 
\end{eqnarray}

A wider discussion on the truncation of the integrals performed along the imaginary axis may be found in \cite{cott09}. 

Now, having represented the transfer function both in exact and approximated form in terms of H-FSM we are ready to infer an analytic expression for the process $Y(t)$ with target PSD.

The input-output relation for linear system in Fourier domain is written as $Y\left( {\omega ,T} \right) = H\left( \omega  \right)W\left( {\omega ,T} \right)$, where $T>0$ is a truncation bound. Of course this relies on the fact that $W(t)$ is a stationary noise.
Thus, bearing in mind eq.(\ref{qq21}) one obtains
\begin{equation}
Y\left( \omega,T  \right) = \frac{1}{{4\pi i}}\int_{\rho  - i\infty }^{\rho  + i\infty } {\Pi _H \left( { - \gamma } \right)\left| \omega  \right|^{\gamma  - 1} W\left( \omega,T  \right){\rm{d}}\gamma } 
\label{qq27}
\end{equation}
due to the linearity of the operator ${\cal{L}}\left(  \bullet  \right)$. Applying an inverse Fourier transform, the response to the linear system assumes the form
\begin{equation}
Y\left( t \right) = \frac{1}{{4\pi i}}\int_{\rho  - i\infty }^{\rho  + i\infty } {\Pi _H \left( { - \gamma } \right){\cal F}^{ - 1} \left\{ {\left| \omega  \right|^{\gamma  - 1} W\left( \omega,T  \right);t} \right\}{\rm{d}}\gamma } 
\label{qq28}
\end{equation}

By recalling eq.(\ref{qq5c}), it follows that 
$\mathop {\lim }\limits_{T \to \infty } {\cal F}^{ - 1} \left\{ {\left| \omega  \right|^{\gamma - 1} W\left( \omega,T  \right);t} \right\} = \left( {I^{1 - \gamma } W} \right)\left( t \right)$, that introduced in the latter, gives 
\begin{equation}
Y\left( t \right) = \frac{1}{{4\pi i}}\int_{\rho  - i\infty }^{\rho  + i\infty } {\Pi _H \left( { - \gamma } \right)\left( {I^{1 - \gamma } W} \right)\left( t \right){\rm{d}}\gamma } 
\label{qq29}
\end{equation}
with $\rho> 0$ as already found by means of the impulse response function, that is the new exact representation of the stationary process with assigned PSD.

Some comments are necessary to highlight the peculiar aspects of this new representation of a stationary process. $i)$ The resulting process $Y(t)$ is reconstructed by the knowledge of the H-FSM previously defined in eqs.(\ref{qq18}), calculated on the transfer function given by eq.(\ref{qq17}), through the target power spectral density $S_Y(\omega)$. There is no need of using optimization criteria like in $ARMA$, or similar models, because the coefficients figuring in the representation have the meaning of being complex moments of the $H(\omega)$. $ii)$ Eq.(\ref{qq29}) is an integral along the imaginary axis with fixed real part $\rho$, belonging to the interval $[0,1]$. The choice of $\rho$ inside this interval does not influence the integral because the integrand is holomorphic inside such a interval. $iii)$ Fractional integrals of white noise processes have recently attracted many authors. In fact, such operators are connected with the so called Fractional Brownian noises and motions as reported in \cite{chec01}, \cite{grig07}. Although white noise processes have nowhere differentiable path, the operation of fractional integration and derivation is indeed meaningful. To get a clue on this, it suffices to recall the definitions in eqs.(\ref{eq4a})-(\ref{eq4b}) which stress the convolution nature of the fractional operators. The kernel of the integral smooths the singularity of the process path in a such a way that it is therefore well-defined. In (\cite{west03}, pp.65-70) very interesting results on a class of nowhere differential function (the Weierstrass function) is also presented. $iv)$ The process $Y(t)$ as expressed in eq.(\ref{qq29}) is suited to be computed by proper discretization of the integral involved. Indeed, truncating the integral up to a certain value, called ${\bar \eta }$, and dividing $\left[ { - \bar \eta ,\bar \eta } \right]$ in $2m$ intervals of amplitude $\Delta \eta  = \bar \eta /m$, the integral can be approximated by the sum
\begin{equation}
Y\left( t \right) = \frac{{\Delta \eta }}{{4\pi }}\sum\limits_{k =  - m}^m {\Pi _H \left( { - \gamma _k } \right)\left( {I^{1 - \gamma _k } W} \right)\left( t \right)} 
\label{qq30}
\end{equation}
with $\gamma _k  = \rho  + ik\Delta \eta$ $\left( {0 < \rho  < 1} \right)$. This approximation carries out a truncation and discretization error, that can be made arbitrarily small. Eq.(\ref{qq29}), or its discretized counterpart given in eq.(\ref{qq30}), are the new representation of the process $Y(t)$ whose PSD matches the target one. In particular eq.(\ref{qq30}) shows that the $Y(t)$ may be obtained as the superposition of the fractional integrals of the white noise process.

In order to validate eq.(\ref{qq30}), we will show that with some simple steps it coincides with the well-known Shinozuka's representation \cite{shin91}. Let us consider a band limited white noise process with one-side PSD
\begin{equation}
G_W \left( \omega  \right) = \left\{ \begin{array}{l}
 q/\pi \,\,\,\,\,\,\,\,0 \le \omega  \le \bar \omega  \\ 
 0\,\,\,\,\,\,\,\,\,\,\,\,\,\,otherwise \\ 
 \end{array} \right.
\label{qq31}
\end{equation}
where ${\bar \omega }$ is some cut-off frequency and let be $\Delta \omega $ a discretization of the $\omega $ axis such that $\Delta \omega  = {{\bar \omega } \mathord{\left/
 {\vphantom {{\bar \omega } n}} \right.
 \kern-\nulldelimiterspace} n}$, then the spectral representation theorem \cite{shin91} 
\begin{equation}
W\left( t \right) =
\mathop {{\rm{lim}}}\limits_{\scriptstyle n \to \infty  \atop 
  \scriptstyle \Delta \omega  \to 0}  
 \sum\limits_{j = 1}^n  {\sqrt {2\Delta \omega \,q/\pi } } \sin \left( {\omega _j t + \phi _j } \right)
\label{qq32}
\end{equation}
holds true, where $\phi _j $ are realizations of a random variable uniformly distributed in $\left[ {0,2\pi } \right]$.  The fractional integral ${\left( {I^{1 - \gamma _k } W} \right)\left( t \right)}$ can be directly computed by Mathematica and is
\begin{equation}
\left( {I^{1 - \gamma _k } W} \right)\left( t \right) = \mathop {{\rm{lim}}}\limits_{\scriptstyle n \to \infty  \atop 
  \scriptstyle \Delta \omega  \to 0} \sum\limits_{j = 1}^n  {\sqrt {2\Delta \omega \,q/\pi } } \left| {\omega _j } \right|^{\gamma_k  - 1} \sin \left( {\omega _j t + \phi _j } \right)
\label{qq33}
\end{equation}

Introduction of the latter in eq.(\ref{qq29}) leads to
\begin{equation}
Y\left( t \right) = \mathop {{\rm{lim}}}\limits_{\scriptstyle n \to \infty  \atop 
  \scriptstyle \Delta \omega  \to 0} \sum\limits_{j = 1}^n  {\sqrt {2\Delta \omega \,q/\pi } } \sin \left( {\omega _j t + \phi _j } \right)\frac{1}{{4\pi i}}\int_{\rho  - i\infty }^{\rho  + i\infty } {\Pi _H \left( { - \gamma } \right)\left| {\omega _j } \right|^{\gamma  - 1} {\rm{d}}\gamma } 
\label{qq34}
\end{equation}
that, bearing in mind eqs.(\ref{qq17}) and (\ref{qq21}), is simplified in the form
\begin{equation}
Y\left( t \right) = \mathop {{\rm{lim}}}\limits_{\scriptstyle n \to \infty  \atop 
  \scriptstyle \Delta \omega  \to 0} \sum\limits_{j = 1}^n  {\sqrt {2G_Y \left( {\omega _j } \right)\Delta \omega } } \sin \left( {\omega _j t + \phi _j } \right)
\label{qq35}
\end{equation}
where $G_Y(\omega)=2 S_Y(\omega)U(\omega)$ is the target unilateral power spectral density. Eq.(\ref{qq35}) is the well known Shinozuka's  representation \cite{shin91} for a process with assigned spectrum, and then the process reconstruction by eq.(\ref{qq30}) is proved.

\subsection{Two Relevant Examples}

\subsubsection{Pierson-Moskowitz spectrum}

Consider the unilateral Pierson-Moskowitz (PM) spectrum \cite{pier69} $G_{PM}$ given by 
\begin{equation}
G_{PM} \left( \omega  \right) = \left\{ {\begin{array}{*{20}c}
   {\frac{{c^{5/4} }}{{\omega ^5 }}\exp \left( { - \frac{5}{{4\omega ^4 }}} \right);} & {\omega  > 0}  \\
   {0;} & {\omega  < 0}  \\
\end{array}} \right.
\label{qq25}
\end{equation}
with assigned parameter $c$. Let us study the linear system excited by a Gaussian white noise process as indicated in eq.(\ref{qq14}) such that $Y(t)$ has a power spectral density with the assigned form $S_Y(\pm \omega)=\frac{1}{2}G_{PM}(\omega)$, with $\omega > 0$. It follows from the definition in eq.(\ref{qq18}) that the H-FSM can be easily evaluated by Mathematica and assume the form
\begin{equation}
\Pi_H(-\gamma)=2^{\frac{1}{8}+\frac{3 \gamma }{4}} 5^{-\frac{3}{8}-\frac{\gamma }{4}} c^{5/8} \sqrt{\pi } \sqrt{\frac{1}{q}} \Gamma\left[\frac{3}{8}+\frac{\gamma }{4}\right]
\label{qq26}
\end{equation}
having introduced $H_{PM}(\omega)=\sqrt {\pi \left| {G_{PM} \left( \omega  \right)} \right|/q} 
$, as given by eq.(\ref{qq16}).

\begin{figure}[ht]
	\centering
		\includegraphics[scale=.20]{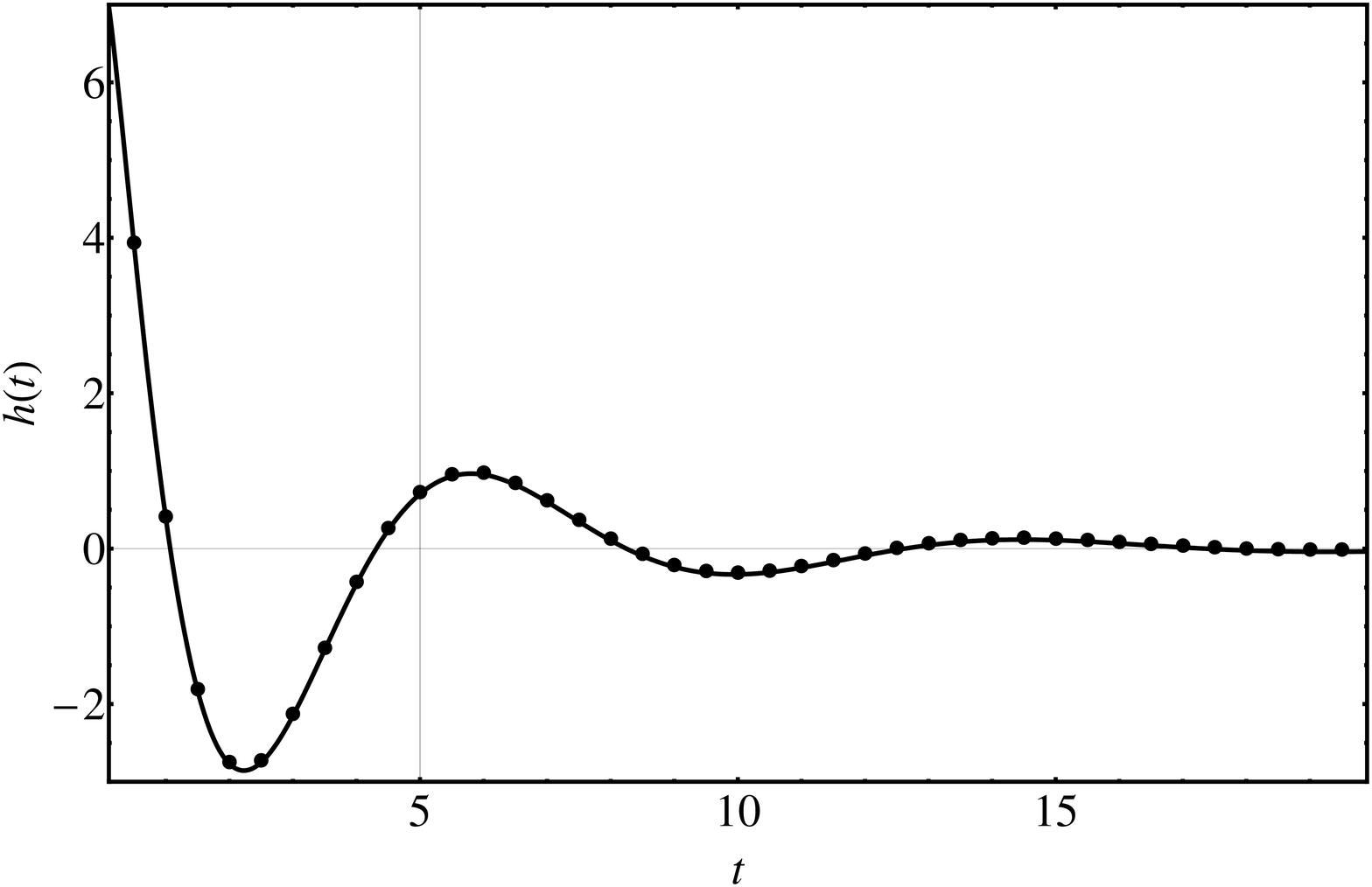}
	\caption{Approximate form of $h_{PM}(t)$ (dotted) contrasted with ${\cal F}^{ - 1} H\left( \omega  \right)$}
	\label{f1}
\end{figure}

\begin{figure}[ht]
	\centering
		\includegraphics[scale=.20]{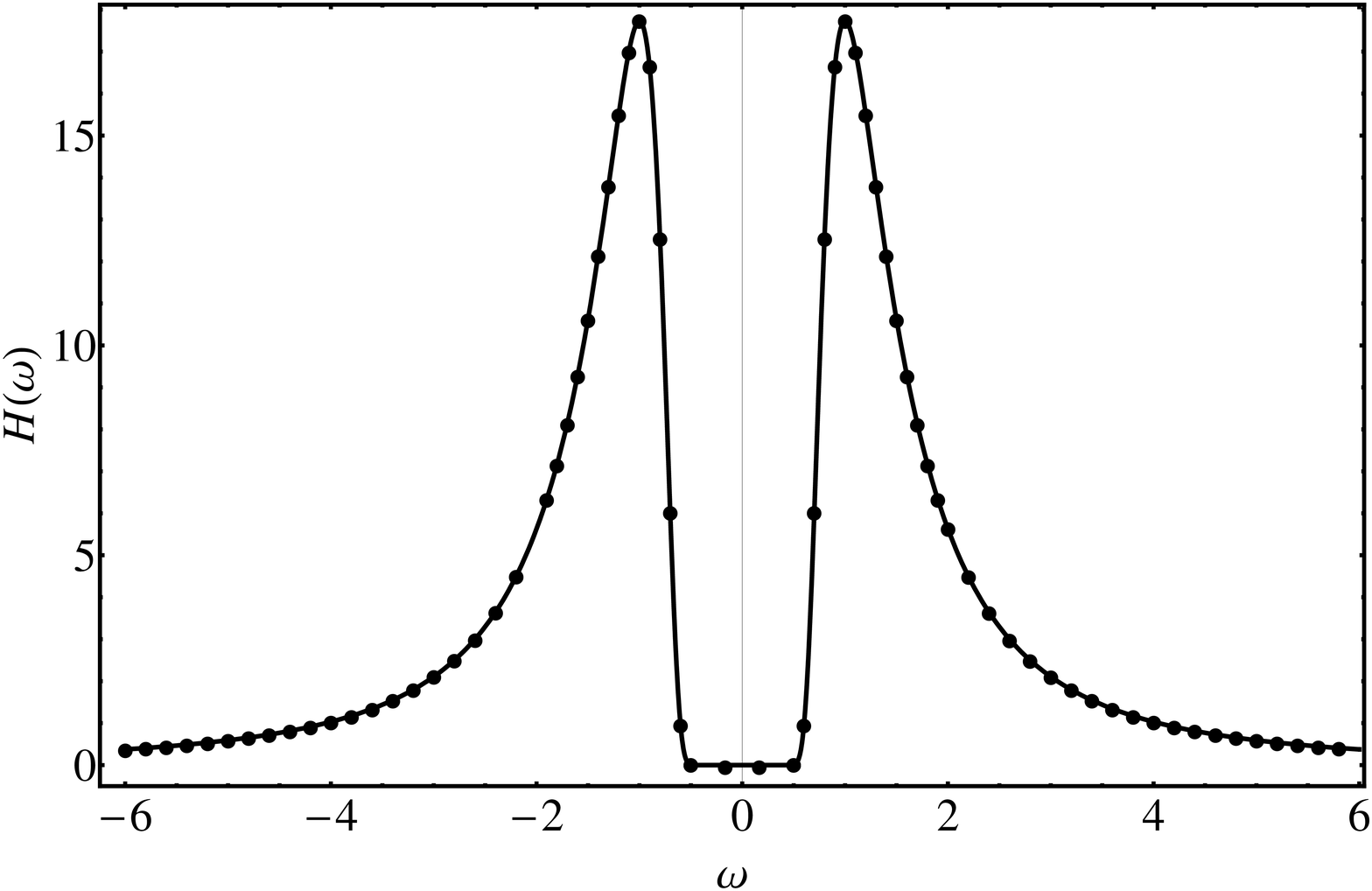}
	\caption{Approximate $H_{PM}(\omega)$ (dotted) contrasted with the exact one}
	\label{f2}
\end{figure}

\begin{figure}[ht]
	\centering
		\includegraphics[scale=.20]{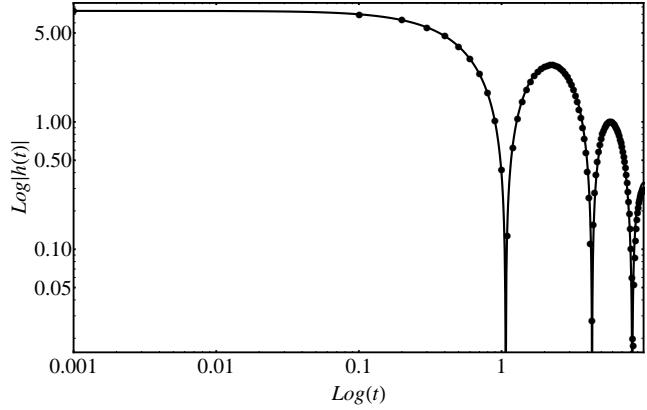}
	\caption{Log-Log plot of the approximate form of $h_{PM}(t)$ (dotted) contrasted with ${\cal F}^{ - 1} H\left( \omega  \right)$}
	\label{f3}
\end{figure}

The approximated impulse response function $h_{PM}(t)$ and the transfer function $H(\omega)$ of the system under exam have been found by applying eqs.(\ref{qq22}) and (\ref{qq23}) and plotted in Fig.(\ref{f1}) and Fig.(\ref{f2}), respectively. The parameters used are: $c=2.72$, $ \rho=0.5$, $m=25$, $\Delta \eta=0.6$ ($ \bar \eta =15 $). It can be noted that the comparison between the exact functions (continuous line) and the approximated one (dotted line) is very good. As confirmed by the log-log plot in Fig.(\ref{f3}), the two functions coincide in a very wide interval. Although in Fig.(\ref{f1}) the impulse response function is plotted only for $t>0$, it is a symmetric function because the differential equation associated to eq.(\ref{qq22}) is non causal.

\subsubsection{Davenport's spectrum}

\begin{figure}[ht]
	\centering
		\includegraphics[scale=.20]{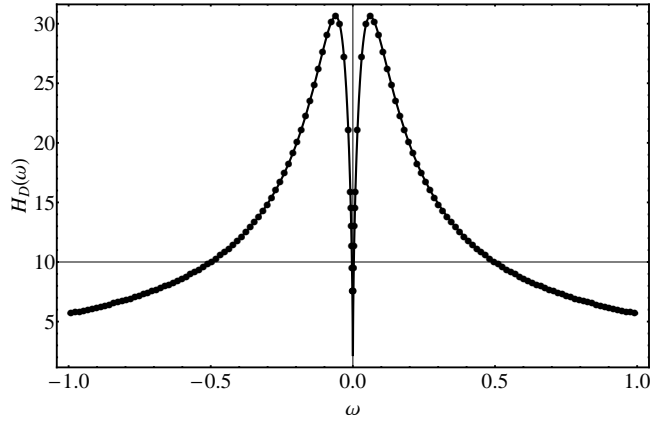}
	\caption{Approximate $H_D(\omega)$ (dotted) contrasted with the exact one}
	\label{f3D}
\end{figure}

In wind engineering, the Davenport's spectrum
\begin{equation}
S_D \left( \omega  \right) = \frac{{4\pi k_0 V_r^2 }}
{{\left| \omega  \right|}}\frac{{q_D\left( \omega  \right)^2 }}
{{\left( {1 + q_D\left( \omega  \right)^2 } \right)^{4/3} }}
\label{eq27}
\end{equation}
is widely used to represent wind fluctuation, where  $V_r$ is the mean wind speed at the reference level, $k_0$ is a roughness characteristic of the analyzed site and $q_D(\omega)=1200\omega/(2\pi V_r)$. For this application we selected $k_0=0.01$, $Vr=15 m/s$, $q=1$  as spectrum parameters and $ \rho=0.7$, $m=30$, $\Delta \eta=0.5$ ($ \bar \eta =15 $). In fig.(\ref{f3D}) the approximated transfer function $H_{D}(\omega)$ is contrasted with the exact one defined from eq.(\ref{eq27}). With this further example we want to stress that the spectrum behavior in the neighborhood of the zero has no influence on the applicability and the efficiency of the method. Indeed both the flat and the steep behavior of the PM and the Davenport spectra, are very good approximated by H-FSM.

\section{Fractional differential equations of the linear filter}

The method presented in the previous section characterizes the process $Y(t)$ with assigned spectrum, by the exact expression in eq.(\ref{qq29}) or by the approximated one in eq.(\ref{qq30}). In this section, the differential equation whose response is $Y(t)$ will be found in approximated form, making extensive use of composition properties of fractional operators reported in Appendix A3. 
Firstly, let us rewrite eq.(\ref{qq30}) in the form $Y\left( t \right) \cong \sum\limits_{k =  - m}^m {Y_k \left( t \right)} $ where
\begin{equation}
Y_k \left( t \right) = \alpha_k \left( {I_{}^{1 - \gamma _k } W} \right)\left( t \right)
\label{qq36}
\end{equation}
with $\alpha_k  = \Delta \eta \Pi _H \left( { - \gamma _k } \right)/\left( {4\pi } \right)$. Applying to both sides of the latter equation the operator $D^{1 - \gamma _k } $ and exploiting the composition rule reported in Appendix A3 leading to eq.(\ref{bq51}), we obtain
\begin{equation}
\left( {D_{}^{1 - \gamma _k } Y_k } \right)\left( t \right) = \alpha_k W\left( t \right)
\label{qq39}
\end{equation}

Each component of $Y(t)$ is therefore the solution of a linear fractional stochastic differential equation excited by a Gaussian white noise. As the noise is the same for each $Y_k (t)$, the k components are dependent each other.
For a complete description of the whole process $Y(t)$, it suffices to perform a time derivative of the first order to eq.(\ref{qq36}) and summing for $k =-m,...,m$ obtaining 
\begin{equation}
\frac{{{\rm d} Y}}{{{\rm d} t}} = \sum\limits_{k =  - m}^m {\alpha _k } \left( {D^{\gamma _k } W} \right)\left( t \right)
\label{qq39bis}
\end{equation}
that is the linear fractional stochastic differential equation searched.

In order to provide the probabilistic characterization of the response process $Y(t)$ let us consider the truncated Fourier transform of eq.(\ref{qq39}) yielding
\begin{equation}
\left| \omega  \right|^{1 - \gamma _k } Y_k \left( {\omega ,T} \right) = \alpha _k W\left( {\omega ,T} \right)
\label{qq40}
\end{equation}
or in terms of PSD
\begin{equation}
S_{Y_k } \left( \omega  \right) = \left| {\alpha _k } \right|^2 \left| \omega  \right|^{2\left( {\gamma _k  - 1} \right)} S_W  = \left| {\alpha _k } \right|^2 \left| \omega  \right|^{2\left( {\gamma _k  - 1} \right)} \left( {{q \mathord{\left/
 {\vphantom {q {2\pi }}} \right.
 \kern-\nulldelimiterspace} {2\pi }}} \right)
\label{qq41}
\end{equation}
while the cross spectral density takes the form
\begin{equation}
S_{Y_k Y_j } \left( \omega  \right) = \alpha_k^{} \alpha_j^* \left| \omega  \right|^{\gamma _k  - 1} \left( {\left| \omega  \right|^{\gamma _j  - 1} } \right)^* \left( {{q \mathord{\left/
 {\vphantom {q {2\pi }}} \right.
 \kern-\nulldelimiterspace} {2\pi }}} \right)
\label{qq42}
\end{equation}
where the symbol * means complex conjugate.

It follows from eq.(\ref{qq42}) that the PSD of the target process $Y(t)$ is written as
\begin{equation}
S_Y \left( \omega  \right) = \sum\limits_{k =  - m}^m {\sum\limits_{j =  - m}^m {S_{Y_k Y_j } \left( \omega  \right)} } 
\label{qq43}
\end{equation}

\begin{figure}[ht]
	\centering
		\includegraphics[scale=1]{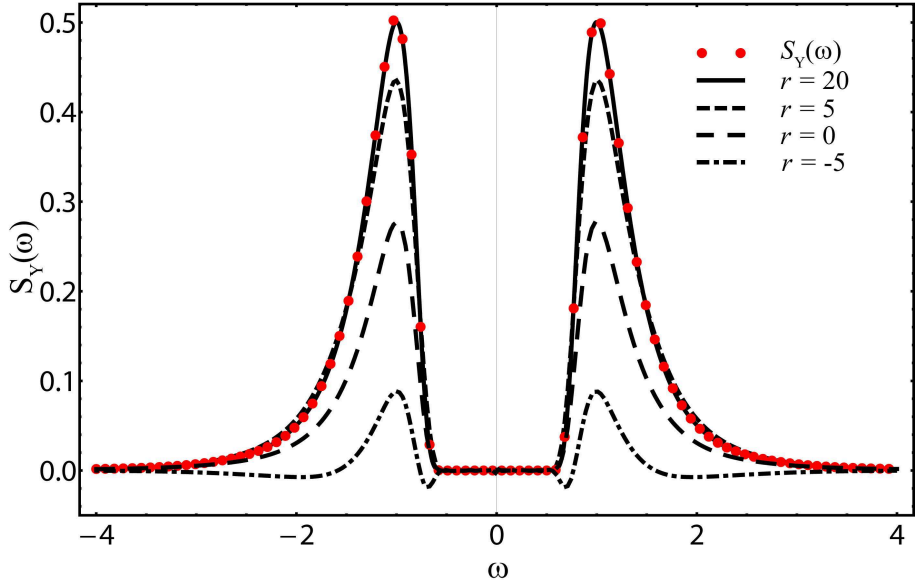}
	\caption{Approximation of the Pierson-Moskowitz spectrum $S_Y(\omega)$, by components' cross-spectral densities $S_Y \left( \omega  \right) = \sum\limits_{k =  - m}^m {\sum\limits_{j =  - m}^r {S_{Y_k Y_j } \left( \omega  \right)} } $, varying $r$
}
	\label{f5}
\end{figure}

Therefore, the cross-spectral density between the dependent components $Y_k$ characterizes the target power spectral density $S_Y(\omega)$. We want to stress that each component $S_{Y_k}$ gives a contribute in defining the total power spectral density $S_Y(\omega)$ and that only the whole ensemble returns a very good approximation. This fact is highlighted in Fig.(\ref{f5}) where the approximation of a Pierson-Moskowitz spectral density, $S_Y(\omega)$, by components' cross-spectral densities $S_Y \left( \omega  \right) = \sum\limits_{k =  - m}^m {\sum\limits_{j =  - m}^r {S_{Y_k Y_j } \left( \omega  \right)} } $, varying $r$ is plotted. This figure shows in particular that every single component $Y_k(t)$ in eq.(\ref{qq36}) gives a contribution, because the convergence is attained only when $r=m$. Summing up, every component $Y_k$ weights in the reconstruction of the target PSD.

\section{Comparison with time series models}

The processes $Y(t)$ and $W(t)$ up to now considered are continuous in time. In this section we show how to express the process $Y(t)$ with target spectral density in a sampled time space. To this aim, let us discretize the time axis in equally spaced intervals of amplitude $\Delta t > 0$. With the notation $Y_{k,t - j} ,{\kern 1pt} j \in \mathbb{N}$ we denote $Y_k \left( {t - j\Delta t} \right)$ and consequently $Y_{k,t}$ is the process at the instant of time $t$. The same notation applies to the noise such that $W_{t - j}  = W\left( {t - j\Delta t} \right)$.

The discretization of the Riesz operator can be tackled by the Gr${\rm \ddot{u}}$nwald-Letnikov
approach as reported in \cite{samk93} that reads
\begin{equation}
D_{}^\beta  Y_{k,t}  = \mathop {\lim }\limits_{\Delta t \to 0} \left( {\sum\limits_{j = 0}^\infty  {\lambda _j \left( \beta  \right)Y_{k,t - j}  + } \sum\limits_{j = 0}^\infty  {\lambda _j \left( \beta  \right)Y_{k,t + j} } } \right)
\label{qq34n}
\end{equation}
with
\begin{equation}
\lambda _j \left( \beta  \right) = \frac{{( - 1)^j \left( {\begin{array}{*{20}c}
   \beta   \\
   j  \\
\end{array}} \right)}}{{2\Delta t^{  \beta } \cos \left( {{{\beta \pi } \mathord{\left/
 {\vphantom {{\beta \pi } 2}} \right.
 \kern-\nulldelimiterspace} 2}} \right)}}
\label{qq35n}
\end{equation}

Eq.(\ref{qq34n}) is very useful in numerical applications because for a small value of $\Delta t > 0$ gives an approximation formula of the first order. For our scope eq.(\ref{qq34n}) is extremely useful because gives the possibility to interpret the fractional filter equation in eq.(\ref{qq39}) also in the discrete time domain in order to make a fruitful comparison with other well-known time series model.

Then, by means of the discretization in step of small $\Delta t > 0$ and substituting the approximated Riesz operator of eq.(\ref{qq34n}) in eq. (\ref{qq39}), the following time series is obtained
\begin{equation}
\sum\limits_{j = 0}^\infty  {\lambda _j \left( {1 - \gamma _k } \right)\left\{ {Y_{k,t - j}  + Y_{k,t + j} } \right\}}  = \alpha_k W_t 
\label{qq36n}
\end{equation}

This representation highlights that this fractional time series model is determined by the knowledge of the past of the process in the first term and the future of the process in the second term. This happens because the filter is non-causal. From eq.(\ref{qq36n}) it is possible to find the transfer function by applying the z-transform. Remanding to the textbook of \cite{oppe75} on details concerning the z-transform, it can be proved that the pulse transfer function for the single component $Y_k$ is
\begin{equation}
H_k \left( z \right) = \frac{{\alpha_k }}{{\sum\limits_{j = 0}^\infty  {\lambda _j \left( {1 - \gamma _k } \right)\left\{ {z^{ - j}  + z^j } \right\}} }} 
\label{qq37n}
\end{equation}

The fractional filter characterized by eq.(\ref{qq36n}) and (\ref{qq37n}) is suitable to be compared with the classical Auto Regressive model of order $p \in \mathbb{N}$, denoted as $AR(p)$, whose time series has the form
\begin{equation}
y_t  + \sum\limits_{j = 1}^p {a_j } y_{t - j}  = GW_t  
\label{qq38n}
\end{equation}
and pulse transfer function given by the equation
\begin{equation}
H_{AR(p)} \left( z \right) = \frac{G}{{1 + \sum\limits_{j = 1}^p {a_j z^{ - j} } }}
\label{qq39n}
\end{equation}

Comparing the pulse transfer function in eq.(\ref{qq37n}) and in eq.(\ref{qq39n}) it is worth to note that both expressions present the $z$ variable only in the denominator, sharing the same algebraic structure. Remarkable difference is that the coefficients in the fractional filter are evaluated by the H-FSM in exact form while in the $AR$ model a Jule-Walker scheme must be adapted for finding the coefficients $a_j$; by contrast once the H-FSM are evaluated, all the coefficients in the fractional differential equation are readily found.

Same kind of analogy can be pursued starting from eq.(\ref{qq36}) that assumes the form
\begin{equation}
Y_{k,t}  = \alpha_k \sum\limits_{j = 0}^\infty  {\lambda _j \left( { - \beta } \right)\left\{ {W_{t - j}  + W_{t + j} } \right\}} 
\label{qq40n}
\end{equation}
in the discrete time model and with transfer function
\begin{equation}
H_k \left( z \right) = \alpha_k \sum\limits_{j = 0}^\infty  {\lambda _j \left( { - \beta } \right)\left\{ {z^{ - j}  + z^j } \right\}} 
\label{qq41n}
\end{equation}
Indeed it is easy to recognize that eq.(\ref{qq41n}) is conceptually similar to a Moving Average model of order $q \in \mathbb{N}$ indicated by $MA(q)$, characterized by the temporal series
\begin{equation}
y_t  = \sum\limits_{j = 0}^q {b_j } W_{t - j}  
\label{qq42n}
\end{equation}
and transfer function
\begin{equation}
H_{MA(q)} \left( z \right) = \sum\limits_{j = 0}^q {b_j } z^{ - j}  
\label{qq43n}
\end{equation}

Comparison between $MA$ model and digital filter obtained by the proposed procedure also in this case reveals the same structure. It is to be stressed that for $MA$ model, the coefficients $b_j$ are evaluated by some optimization criteria while in eq.(\ref{qq41n}) the coefficients $\alpha_k$ are evaluated by H-FSM.

The last step we develop regards the $ARMA$ model and follows plainly. First of all, let us exploit the composition rule given in Appendix in eq.(\ref{qq5b}) into eq.(\ref{qq39}), applying ${D^{\gamma _s  - 1} }$ to both sides of eq.(\ref{qq39}). In this way the filter equation reads
\begin{equation}
\left( {D_{}^{\gamma _s  - \gamma _k } Y} \right)\left( t \right) = \alpha_k \left( {D_{}^{\gamma _s  - 1} W} \right)\left( t \right)
\label{qq44n}
\end{equation}
that in the discrete time form is
\begin{equation}
\sum\limits_{j = 0}^\infty  {\lambda _j \left( {\gamma _s  - \gamma _k } \right)\left\{ {Y_{k,t - j}  + Y_{k,t + j} } \right\}}  = \alpha_k \sum\limits_{j = 0}^\infty  {\lambda _j \left( {1 - \gamma _s } \right)\left\{ {W_{t - j}  + W_{t + j} } \right\}} 
\label{qq45n}
\end{equation}
and whose transfer function is
\begin{equation}
H_k \left( z \right) = \frac{{\alpha_k \sum\limits_{j = 0}^\infty  {\lambda _j \left( {1 - \gamma _s } \right)\left\{ {z^{ - j}  + z^j } \right\}} }}{{\sum\limits_{j = 0}^\infty  {\lambda _j \left( {\gamma _s  - \gamma _k } \right)\left\{ {z^{ - j}  + z^j } \right\}} }}
\label{qq46n}
\end{equation}
The latter should be compared with the $ARMA$ model, that is characterized by the time series model
\begin{equation}
\sum\limits_{j = 0}^p {d _j Y_{t - j}  = } \sum\limits_{j = 0}^q {c_j W_{t - j} } 
\label{qq47n}
\end{equation}
and transfer function 
\begin{equation}
H_{ARMA} \left( z \right) = {{\sum\limits_{j = 0}^q {c_j z^{ - j} } } \mathord{\left/
 {\vphantom {{\sum\limits_{j = 0}^q {c_j z^{ - j} } } {\sum\limits_{j = 0}^p {{\mathop{\rm d}\nolimits} _j z^{ - j} } }}} \right.
 \kern-\nulldelimiterspace} {\sum\limits_{j = 0}^p {d _j z^{ - j} } }}
\label{qq48n}
\end{equation}

Also in this case direct comparison between eq.(\ref{qq46n}) and eq.(\ref{qq48n}) shows the correspondence between the two representations.

\section{Conclusions}

This paper sheds a new light in the representation of stationary Gaussian colored noises. It is shown that every stationary Gaussian process with assigned power spectral density is equal to an integral involving the H-Fractional Spectral Moments (H-FSM) and Riesz Fractional integrals of the Gaussian white noise process. In this sense, the first result of this paper is a new exact representation for stationary processes.

It has been further shown that the H-FSM can be easily evaluated, and two pathological applications like the Pierson-Moskowitz and the Davenport density spectra have been reported. 

Moreover, properly approximating the exact integral representation in to a series form, a set of fractional stochastic differential equations excited by an external white noise process has been provided. Its solution has the desired target power spectral density. It has also been shown that every stationary Gaussian process is approximated by a fractional stochastic differential equations, that is a stochastic differential equation excited by fractional derivatives of a Gaussian white noise.

The representation proposed has been rewritten in time series form and compared with $AR$, $MA$ and $ARMA$ time series.

\appendix
\section{Appendix}\nonumber

The generalized Taylor form applied in the paper to symmetric functions expressed in terms of Riesz operators is valid also once this condition is dropped, as reported in \cite{cott09}. For readability's sake we report the main results, introducing the proper operator of Riemann-Liouville and Marchaud.

\subsection{Riemann-Liouville (RL) and Marchaud fractional operators}

We recall the definitions of the Riemann-Liouville (RL) fractional integral and derivative, $\left( {I_ \pm ^\gamma  f} \right)\left( t \right)$ and $\left( {{\cal D}_ \pm ^\gamma  f} \right)\left( t \right)$, respectively, and of the Marchaud fractional derivative, $\left( {{\bm{D}}_ \pm ^\gamma  f} \right)\left( t \right)$, given by
\begin{subequations}\label{eq3}
\begin{eqnarray}\label{eq3a}
\left( {I_ \pm ^\gamma  f} \right)\left( t \right)\mathop  = \limits^{def} \frac{1}{{\Gamma \left( \gamma  \right)}}\int_0^\infty  {\xi ^{\gamma  - 1} f\left( {t \mp \xi } \right)\rm{d} \xi } 
\end{eqnarray}
\begin{eqnarray}\label{eq3b}
\left( {{\cal D}_ \pm ^\gamma  f} \right)\left( t \right)\mathop  = \limits^{def} \frac{{ \pm 1}}{{\Gamma \left( {1 - \gamma } \right)}}\frac{{\rm d} }{{{\rm d} t}}\int_0^\infty  {\xi ^{ - \gamma } f\left( {t \mp \xi } \right){\rm d} \xi } 
\end{eqnarray}
\begin{eqnarray}\label{eq3c}
\left( {{\bm{D}}_ \pm ^\gamma  f} \right)\left( t \right)\mathop  = \limits^{def} \frac{1}{{\Gamma \left( { - \gamma } \right)}}\int_0^\infty  {\xi ^{ - \gamma  - 1} \left( {f\left( {t \mp \xi } \right) - f\left( t \right)} \right){\rm{d}}\xi } 
\end{eqnarray}
\end{subequations}
where $\Gamma \left(  \bullet  \right)$ is the Euler gamma function and $\gamma = \rho +\rm{i}\eta, \rho >0, \eta \in \mathbb{R}$. 
The Marchaud definition of the fractional derivatives is more convenient with respect to the Riemann-Liouville in the sense that they exist also for function that do not vanish at infinity, and for this reason it will be preferred.

Fourier transforms of the fractional derivatives and integrals above defined are
\begin{subequations}\label{qq5n}
\begin{eqnarray}\label{qq5a}
{\cal F}\left\{ {\left( {I_ \pm ^\gamma  f} \right)\left( t \right);\omega } \right\} = \left( { \mp {\rm{i}}\omega } \right)^{ - \gamma } {\cal F}\left\{ {f\left( t \right);\omega } \right\}
\end{eqnarray}
\begin{eqnarray}\label{qq5b}
{\cal F}\left\{ {\left( {{\bm{D}}_ \pm ^\gamma  f} \right)\left( t \right);\omega } \right\} = \left( { \mp {\rm{i}}\omega } \right)^\gamma  {\cal F}\left\{ {f\left( t \right);\omega } \right\}
\end{eqnarray}
\end{subequations}

Applying same reasoning reported in section 1 for the Riesz operators, comparing eqs.(\ref{qq5a})-(\ref{qq5b}) it can be simply verified that for functions that are Fourier transformable $\left( {\bm{D}_ \pm ^\gamma  f} \right)\left( t \right) \equiv \left( {I_ \pm ^{-\gamma}  f} \right)\left( t \right)$, although not valid in general. Then, by making an inverse Fourier transform of the eq.(\ref{qq5n}) the following representations
\begin{equation}\label{qq6an}
\left( {I_ \pm ^\gamma  f} \right)\left( t \right) = ({\bm{D}}_\pm^{-\gamma} f)(t)=\frac{1}{{2\pi }}\int_{ - \infty }^\infty  {\left( { \mp i\omega } \right)^{ - \gamma } \varphi \left( \omega  \right)e^{ - i\omega t} {\rm {d}}\omega } 
\end{equation}
holds true. Moreover, eq.(\ref{qq6an}) evaluated in $t=0$ assumes the particular meaning of fractional moments of $\varphi(\omega)$ in the form
\begin{equation}
\mu _ \mp  \left( { - \gamma } \right)\mathop  = \limits^{def} 2\pi \left( {I_ \pm ^\gamma  f} \right)\left( 0 \right) = \int_{ - \infty }^\infty  {\left( { \mp i\omega } \right)^{ - \gamma } \varphi \left( \omega  \right){\rm d}\omega } 
\label{qq7}
\end{equation}

Once the fractional moments have been introduced, let us interpret the fractional operators in eq.(\ref{qq7}) as Mellin transform of the function $f(t)$
\begin{equation}
\left( {I_ \pm ^\gamma  f} \right)\left( 0 \right)\Gamma \left( \gamma  \right) = \int_0^\infty  {\xi ^{\gamma  - 1} f\left( { \mp \xi } \right)} {\mathop{\rm d}\nolimits} \xi 
\label{qq9}
\end{equation}
as suggested in \cite{samk93}. Then, making an inverse Mellin transform, the following representation of the function $f(t)$
\begin{equation}
f\left( { \mp t } \right) = \frac{1}{{2\pi i}}\int_{\rho  - i\infty }^{\rho  + i\infty } {\Gamma \left( \gamma  \right)\left( {I_ \pm ^\gamma  f} \right)\left( 0 \right) t ^{ - \gamma } {\rm d} \gamma } 
\label{qq10}
\end{equation}
holds true, where the integral is performed along the imaginary axis with fixed real part $\rho$ where $\rho$ belongs to the \textit{fundamental strip} of the Mellin transform. 

By inserting eq.(\ref{qq7}) in eq.(\ref{qq10}), the integral representation for both $f(t)$ and its Fourier transform may be written as 
\begin{subequations}\label{qq11}
\begin{eqnarray}\label{qq11a}
  f\left( { \mp t} \right) = \frac{1}{{2\pi i}}\int_{\rho  - i\infty }^{\rho  + i\infty } {\frac{{\Gamma \left( \gamma  \right)}}{{2\pi }}\mu _ \mp  } \left( { - \gamma } \right)t^{ - \gamma } {\mathop{\rm d}\nolimits} \gamma 
\end{eqnarray}
\begin{eqnarray}
  \varphi \left( \omega  \right) = \frac{1}{{\left( {2\pi } \right)^2 i}}\int_{\rho  - i\infty }^{\rho  + i\infty } {\Gamma \left( \gamma  \right)\Gamma \left( {\gamma  - 1} \right)\left[ {\mu _ -  \left( { - \gamma } \right)\left( {i\omega } \right)^{\gamma  - 1} {\rm{ + }}\mu _ +  \left( { - \gamma } \right)\left( { - i\omega } \right)^{\gamma  - 1} } \right]{\rm{d}}\gamma } 
\label{qq11b}
\end{eqnarray}
\end{subequations}

\subsection{Some concepts on Mellin transform}

The Mellin transform (\cite{samk93}, p.25; \cite{sned51}, p.41) of a function $f\left( x \right)$, $x > 0$, $s \in \mathbb{C}$ is defined as 
\begin{equation}
\varphi \left( s \right) = \,{\cal M}\left\{ {f\left( x \right);s} \right\} = \int _0^\infty  f\left( \xi  \right)\,\xi ^{\,s{\rm{ - 1}}} d\xi 
\label{aq44}
\end{equation}
along the inverse operator 
\begin{equation}
f\left( x \right) = \,{\cal M}^{\, - {\rm{1}}} \left\{ {\varphi \left( s \right);x} \right\} = \left( {2\pi i} \right)^{ - 1} \int _{c - i\infty }^{c + i\infty } \varphi \left( s \right)\,x^{ - s} ds \label{aq45}
\end{equation}
with $c = {\rm Re} \left( s \right) $

From the definition, it follows that the convergence of the Mellin transform depends on the behavior of the function $f(x)$ at zero and infinity, that is, assuming 
\begin{equation}
f\left( x \right) = \left\{ {\begin{array}{*{20}c}
   {O\left( {x^p } \right),\,\,x \to 0}  \\
   {O\left( {x^q } \right),\,\,x \to \infty }  \\
\end{array}} \right.
\label{aq46}
\end{equation}
the complex function $\varphi \left( s \right)$ is analytic inside the so called \textsl{fundamental strip} $ - p <{\rm Re} s\ <  - q$. The inverse Mellin transform must be calculated selecting $c = {\mathop{\rm Re}\nolimits} s$ inside the fundamental strip. Tables of Mellin transforms of commonly used functions are given in (\cite{sned51}, p.527), while for a complete theory readers are referred to \cite{mari82}.

\subsection{Compositions rules for Riesz fractional derivatives}

By manipulating the definition of the Riesz fractional operators, it can be proved that the following identities
\begin{subequations}\label{bq43}
\begin{eqnarray}\label{bq43a}
\left( {I_{}^\beta  f} \right)\left( t \right) = \frac{1}
{{2 \rm{cos(\beta \pi/2 )}}}\left\{ {\left( {I_ + ^\beta  f} \right)\left( t \right) + \left( {I_ - ^\beta  f} \right)\left( t \right)} \right\}
\end{eqnarray}
\begin{eqnarray}\label{bq43b}
\left( {{\bm{D}}_{}^\beta  f} \right)\left( t \right) = \frac{1}{{2\cos \left( {\beta \pi /2} \right)}}\left\{ {\left( {{\bm{D}}_ + ^\beta  f} \right)\left( t \right) + \left( {{\bm{D}}_ - ^\beta  f} \right)\left( t \right)} \right\}
\end{eqnarray}
\end{subequations}
in terms of RL fractional integrals and Marchaud derivatives hold true.

The left handed ${\left( {I_ + ^\beta  f} \right)\left( t \right)}$ and the right handed ${\left( {I_ - ^\beta  f} \right)\left( t \right)}$ operators, are connected by the equations
\begin{subequations}\label{bq44}
\begin{eqnarray}\label{bq44a}
\left( {I_ - ^\beta  f} \right)\left( t \right) = \cos \left( {\beta \pi } \right)\left( {I_ + ^\beta  f} \right)\left( t \right) + \sin \left( {\beta \pi } \right){\cal H}\left\{ {\left( {I_ + ^\beta  f} \right)\left( s \right);t} \right\}
\end{eqnarray}
\begin{eqnarray}\label{bq44b}
\left( {I_ + ^\beta  f} \right)\left( t \right) = \cos \left( {\beta \pi } \right)\left( {I_ - ^\beta  f} \right)\left( t \right) - \sin \left( {\beta \pi } \right){\cal H}\left\{ {\left( {I_ - ^\beta  f} \right)\left( s \right);t} \right\}
\end{eqnarray}
\end{subequations}
where  $\mathcal{H}$ is the Hilbert transform defined as 
\begin{equation}
{\cal H}\left\{ {f\left( s \right);t} \right\} = \frac{1}{\pi }\int_{ - \infty }^\infty  {\frac{{f\left( s \right)}}{{t - s}}{\rm d} s} 
\label{bq45}
\end{equation}

The eqs.(\ref{bq44}) are useful to find the so called \textsl{composition rules} of fractional operators, which allow to simplify forms of the type ${\bm{D}}_ \pm ^\alpha  I_ \mp ^\beta  f$. In particular, applying ${{\bm{D}}_ + ^\beta  }$ and ${{\bm{D}}_ - ^\beta  }$, to eq.(\ref{bq44a}) and eq.(\ref{bq44b}), respectively, and taking into account  
\begin{equation}
\left( {{\bm{D}}_ + ^\beta  I_ + ^\beta  f} \right)\left( t \right) = f\left( t \right)\,\,\,\,\,\,\,\,\,\,\,\,\,\,\,\,\,\,\,\,\,\,\,\left( {{\bm{D}}_ - ^\beta  I_ - ^\beta  f} \right)\left( t \right) = f\left( t \right)
\label{bq46}
\end{equation}
it follows
\begin{subequations}\label{bq47}
\begin{eqnarray}\label{bq47a}
 \left( {{\bm{D}}_ + ^\beta  I_ - ^\beta  f} \right)\left( t \right) = \cos \left( {\beta \pi } \right)f\left( t \right) + \sin \left( {\beta \pi } \right){\cal H}\left\{ {f\left( s \right);t} \right\}
\end{eqnarray}
\begin{eqnarray}\label{bq47b}
 \left( {{\bm{D}}_ - ^\beta  I_ + ^\beta  f} \right)\left( t \right) = \cos \left( {\beta \pi } \right)f\left( t \right) - \sin \left( {\beta \pi } \right){\cal H}\left\{ {f\left( s \right);t} \right\}
\end{eqnarray}
\end{subequations}
having used the property
\begin{equation}
{\cal H}\left( {I_ \pm ^\beta  f} \right)\left( t \right) = \left( {I_ \pm ^\beta  {\cal H}f} \right)\left( t \right)
\label{bq48}
\end{equation}

Composition rules involving the Riesz operators can now be worked out based on the previous properties, showing that $D^\beta  I^\beta  f = f$. Indeed, from the definitions in eq.(\ref{bq43}) 
\begin{equation}
D^\beta  I^\beta  f = \frac{1}{{\left( {2\cos \left( {\beta \pi /2} \right)} \right)^2 }}\left( {D_ + ^\beta  I_ + ^\beta  f + D_ - ^\beta  I_ - ^\beta  f + D_ + ^\beta  I_ - ^\beta  f + D_ - ^\beta  I_ + ^\beta  f} \right)
\label{bq49}
\end{equation}
that using eq.(\ref{bq46}), (\ref{bq47}) and
\begin{equation}
\frac{{2 + 2\cos \left( {\beta \pi } \right)}}{{\left( {2\cos \left( {\beta \pi /2} \right)} \right)^2 }} = 1
\label{bq50}
\end{equation}
gives the result searched
\begin{equation}
D^\beta  I^\beta  f = f
\label{bq51}
\end{equation}

Other compositions rules and applicability criteria of the formulas in this appendix are adapted from (\cite{samk93}, Chapter 3).

\bibliographystyle{plain}
\bibliography{biblio}

\end{document}